\documentclass[aps,prb,twocolumn,superscriptaddress,showpacs,citeautoscript]{revtex4-1}

\usepackage{amsmath,amssymb}
\usepackage{bm}
\usepackage{graphicx}
\usepackage{epstopdf}
\usepackage{latexsym}
\usepackage{subfigure}
\usepackage{color}
\usepackage{dsfont} 
\usepackage{wasysym} 

\newcommand{\nio}{Na$_4$Ir$_3$O$_8$}

\begin{document}

\title{Anomalous enhancement of Wilson ratio in a quantum spin liquid with strong spin-orbital entanglement: the case of Na$_4$Ir$_3$O$_8$}

\author{Gang Chen}
\email{email address: gang.chen@colorado.edu}
\affiliation{Department of Physics, University of Colorado, Boulder, CO, 80309-0390, U.S.A.}
\author{Yong Baek Kim}
\affiliation{Department of Physics, University of Toronto, Toronto, Ontario M5S 1A7, Canada}
\affiliation{School of Physics, Korea Institute for Advanced Study, Seoul 130-722, Korea}

\begin{abstract}
We present a theory for the metal-insulator transition (MIT) in the quantum-spin-liquid candidate material Na$_4$Ir$_3$O$_8$. We consider an extended Hubbard model on the hyperkagome lattice, which incorporates atomic spin-orbit coupling (SOC) and multi-orbital interactions of iridium 5$d$ electrons. This model is analyzed using the slave-rotor mean-field theory and thermodynamic properties across the MIT are studied. The ground state in the insulating side is a U(1) quantum spin liquid with spinon Fermi surfaces that consist of multiple particle-like and hole-like pockets. It is shown that the Wilson ratio in the quantum spin liquid phase is highly enhanced compared to the metallic state. This originates from the fact that the magnetic susceptibility in the quantum spin liquid phase acquires multiple enhancements due to the strong SOC, reduced band-width and on-site spin-orbital exchange, while the heat capacity does not change much across MIT. This explains the large Wilson ratio of the insulating phase observed in the previous experiment on Na$_4$Ir$_3$O$_8$. 
Possible connections to other existing and future experiments, in particular on the metallic phase, are discussed.
\end{abstract}

\date{\today}


\maketitle

\section{Introduction}

Quantum spin liquid (QSL) is an exotic state of matter with fractionalized elementary excitations 
and emergent gauge fields. Theories of QSLs are well established, but 
it has been extremely challenging to identify a direct evidence for the 
existence of QSLs in real materials.
Recently, however, several materials have been proposed as promising candidates of
QSLs\cite{Balents10}. Magnetic ordering has not been observed in these systems
down to very low temperatures and there exist thermodynamic signatures indicating 
the presence of gapless spin-carrying excitations while there exists a finite charge excitation gap. 
One natural explanation of these low energy spin-carrying excitations would be 
the existence of the fractionalized spinons of QSL phases. The central question in the 
context of QSLs, therefore, is to understand what types of quantum spin liquid phases 
may be realized in these systems. 

In this work, we present a theory of a three-dimensional QSL phase that
may be realized in Na$_4$Ir$_3$O$_8$, an Ir-based hyperkagome lattice system.
In particular, it is shown that a U(1) quantum spin liquid with Fermi surfaces 
and strong spin-orbital entanglement, that results from the large spin-orbit coupling of the Ir 5$d$ electrons
and multi-orbital interactions,
can explain the anomalously large Wilson ratio and Fermi-liquid-like
thermodynamic properties, observed in the insulating phase of this material. 

In the previous experiment on polycrystalline samples of \nio\, in Ref.~\onlinecite{PhysRevLett.99.137207}, 
the magnetic susceptibility and heat capacity 
do not show any signature of magnetic ordering down to very low temperatures.
NMR Knight shift measurement further confirms the absence of magnetic ordering 
down to 2K\cite{Perry}. 
With the Curie-Weiss tempeature $\Theta_{CW} = -650$K, the frustration parameter 
(defined as $f \equiv \Theta_{CW}/{T_N}$ with $T_N$ the ordering temperature) 
is greater than 300. Moreover, in the zero temperature limit, 
the magnetic susceptibility saturates to a large and finite constant value while the heat capacity 
develops a linear temperature dependence with a rather small coefficient $\gamma$
 ($\gamma \equiv {C_v}/{T}$). This leads to the anomalously large Wilson ratio 
$W \equiv ({\pi^2}/{3}) (\chi/\mu_B^2)/({\gamma/k_B^2}) \sim 35$ 
at low temperatures.
Most of other promising QSL candidate materials
have the Wilson ratio of order of unity at low temperatures \cite{Balents10}.
In particular, the organic materials such as $\kappa$-(BEDT-TTF)$_2$Cu$_2$(CN)$_3$\cite{PhysRevLett.95.177001}
and EtMe$_3$Sb[Pd(dmit)$_2$]$_2$\cite{PhysRevB.77.104413} 
also show constant magnetic susceptibility and linear-in-temperature heat capacity
that are often taken as evidences for the spinon Fermi surface.
These two materials are close to a MIT and it has been suggested that 
a QSL phase of spinon Fermi surfaces may arise upon gapping out charge 
excitations via a second order phase transition starting from the metallic side\cite{PhysRevB.78.045109,PhysRevLett.102.186401}. 
In this case, the spinon Fermi surface would be a remnant of the electron Fermi 
surface and the Wilson ratio would not change much across MIT.
While \nio\, may also be close to a MIT due to the extended
nature of Ir $d$-orbitals, a simple extension of this argument would not explain the
large Wilson ratio. Clearly, a different or an additional physics is at work for \nio.

Recently there have been new experiments on single crystal samples\cite{Perry}. 
Depending on the preparation, the single crystal samples can be both insulating and metallic. 
Comparing the conducting metallic samples and the insulating samples, 
one finds that the heat capacity variation is rather small
while the magnetic susceptibility is greatly increased in the insulating samples, leading to
the anomalously large Wilson ratio,
just like in the previous experiment on the insulating polycrystalline sample.
While the origin of the metallic and insulating behaviors in single crystals
is currently not clear, the greatly enhanced susceptibility in the insulating phase 
is clearly the general trend.

There exist several theoretical proposals about the nature of the disordered state 
in the insulating side\cite{PhysRevLett.100.227201,PhysRevLett.101.197201,PhysRevLett.101.197202,PhysRevLett.105.237202}. 
It should be pointed out that all of the existing proposals have assumed a 
spin-$1/2$ Heisenberg model on a hyperkagome lattice. 
Ref.~\onlinecite{PhysRevLett.101.197202} suggested a U(1) QSL with spinon
Fermi surfaces based on a projective wavefunction study. This QSL state is certainly qualitatively 
consistent with the Fermi-liquid-like phenomenology in \nio. The Wilson ratio in this proposal, however, should still be order of unity because of the spin-rotational invariance. 
Ref.~\onlinecite{PhysRevLett.101.197201} introduced a Z$_2$ QSL with spinon pairing. 
They attribute the large Wislon ratio to the suppression of the heat capacity by the spinon
pairing. Even though the singlet spinon pairing suppresses the magnetic susceptibility, 
it is argued that this can be avoided if the SOC energy scale is greater than the pairing strength. 
This state, if relevant for \nio, may lead to emergent superconductivity for the 
metallic samples at similar temperature scales. If there is no superconductivity, 
the metallic samples are expected to have much larger heat capacity compared to 
the insulating ones. Neither seems to be observed in the recent experiments\cite{Perry}.  

Although the investigation of the ground state of the Heisenberg model on the hyperkagome lattice itself 
is an interesting theoretical problem, the applicability of this model to the actual material 
can only be justified in the strong Mott regime \cite{PhysRevB.78.094403}. 
On the other hand, both the polycrystalline sample and single crystal samples are 
proximate to a MIT.
We also notice that all other QSL candidate materials mentioned earlier can be modeled by either a 
single-band Hubbard model in the intermediate or strong Mott regime.
On the other hand, in \nio\,, all the three $t_{2g}$ orbitals are involved and 
hence the electron interactions are dependent on orbital degrees of freedom. 
Moreover, since Ir is a heavy element, a large SOC is naturally expected\cite{PhysRevB.78.094403}. 
An important question is which aspects of these ingredients give rise to qualitatively 
different thermodynamic behaviors. 

In this paper, we analyze the extended Hubbard model 
on the Ir-based hyperkagome lattice, including all of three t$_{2g}$ orbitals, 
SOC, and the electron interactions on these orbitals.   
We assume that the MIT is controlled by the correlation.
We separate the multi-orbital interactions into the Hubbard-$U$ interaction 
for the charge sector and the on-site spin-orbital exchange for the spin-orbital sector. 
We study the MIT in this model by the slave-rotor mean-field theory.
The main result of this paper is that, both of the strong SOC and the correlation 
suppresses the electron bandwidth, which effectively enhances the on-site 
spin-orbital exchange in the insulating side. SOC breaks the spin-rotational 
symmetry and enhances the bare electron/spinon 
magnetic susceptibility, and the magnetic susceptibility is further 
enhanced by the ``enhanced'' on-site spin-orbital exchange interaction. 
While the magnetic susceptibility acquires multiple enhancements, 
the heat capacity is only sensitive to the density of state on the Fermi surface
which does not experience strong enhancement. As a result,
this leads to a large enhancement in the Wilson ratio across the MIT. 
The rest of the paper is organized as follows. 
In Sec.~II, we introduce an extended Hubbard model for Na$_4$Ir$_3$O$_8$.
This model with only the charge sector is solved in the slave-rotor mean-field theory 
in Sec.~III. The spin-orbital exchange interaction is introduced and shown
to be important for thermodynamic properties across the MIT in Sec.~IV.
We conclude and discuss further implications in Sec.~V.

\section{Extended Hubbard Model}
To write down our model, we make use of a few well-known facts about the 
microscopic details related to \nio\cite{PhysRevB.78.094403}.
The local IrO$_6$ crystal field first splits the 5$d$ electron states into 
a low-lying t$_{2g}$ triplet and a much higher-energy e$_g$ doublet that 
will play no role. The atomic SOC acts on the three t$_{2g}$ orbitals. 
Moreover, there may also exist further crystal field splitting among the three
t$_{2g}$ orbitals, which arises from the small distortion of the local IrO$_6$ octahedron\cite{PhysRevLett.99.137207}.
Including the hopping processes of electrons, 
the non-interacting part of the Hamiltonian can be written as
\begin{eqnarray}
{\mathcal H}_0 &=& \sum_{i,mn,\alpha\beta} \frac{\lambda}{2}   {\bf L}_{mn} 
\cdot \boldsymbol{\sigma}_{\alpha\beta}  d^{\dagger}_{im\alpha} d^{\phantom{\dagger}}_{in\beta} 
+ \sum_{i,m,\alpha} D_{im}   d^{\dagger}_{im\alpha} d^{\phantom{\dagger}}_{im\alpha}
 \nonumber 
\\
 &+& \sum_{\langle ij \rangle, mn,\alpha} t_{ij,mn} d^{\dagger}_{im\alpha} d^{\phantom{\dagger}}_{jn\alpha}.
\label{eq:eq1}
\end{eqnarray}
Here $d^{\phantom{\dagger}}_{im\alpha}$ describes the electron with orbital $m=xy,yz,xz$ and
spin $\alpha=\uparrow,\downarrow$ at site $i$, and ${\bf L}$ ($\boldsymbol{\sigma}$) is 
the orbital angular momentum (spin Pauli matrix).
In Eq.~\eqref{eq:eq1}, the single-ion anisotropy $D_{im}$ comes from the distortion 
of the IrO$_6$ octahedra and is different for different Ir sublattices. 
For the hopping amplitudes, we include both the direct and oxygen $p$-orbital mediated 
indirect processes.

\begin{figure}[htp]
\includegraphics[width=6cm]{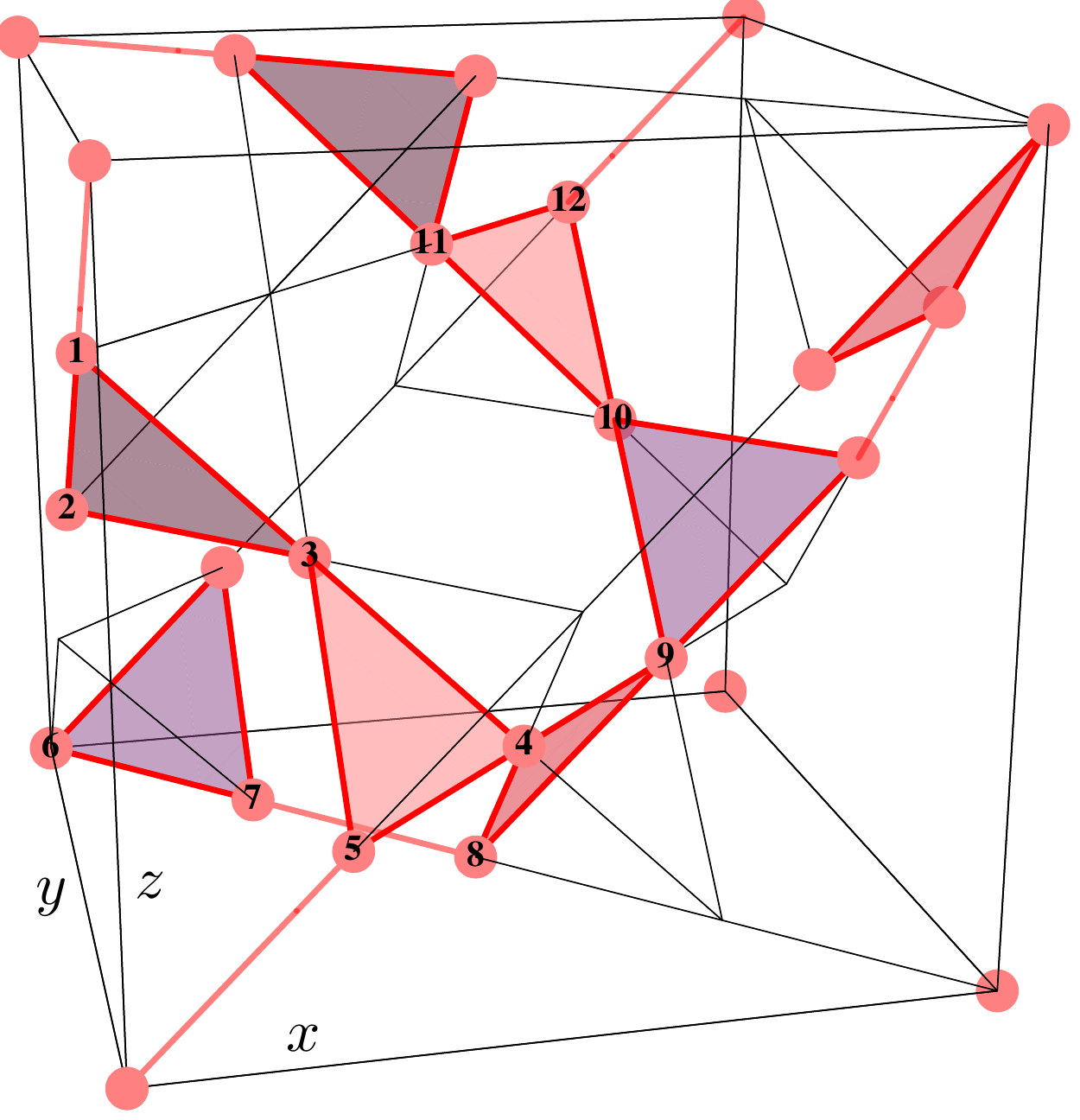}
\caption{(Color online) Ir-based hyperkagome lattice and its parent pyrochlore lattice. 
``1,2,$\cdots$,12'' label the 12 sublattices. Solid bonds (in red) connect Ir sites. }
\label{fig:fig4}
\end{figure}

Now we explain different terms of Eq.~\eqref{eq:eq1} in more detail. 
As shown in Fig.~\ref{fig:fig4}, the Ir hyperkagome lattice has 12 sublattices in the unit cell. 
At each Ir site, there exists a 2-fold symmetry axis\cite{PhysRevB.78.094403} and
the symmetry-allowed single-ion anisotropy has several independent terms. 
We keep the term that is likely to be the dominant single-ion anisotropy. 
For instance, the anisotropy at the sublattice 4 in Fig.~\ref{fig:fig4} is given by 
$\sum_{mn,\sigma} D [(l^z)^2]_{mn} d^{\dagger}_{im\sigma} d^{\phantom{\dagger}}_{in\sigma}$,
where ${\bf l}_{mn} = {\bf L}_{mn}$ and $D>0$. 
Anisotropies on other sublattices can be readily obtained by space group symmetry operations. 

Our tight-binding model contains three parameters: $\sigma$-bonding $t_{\sigma}$ and
$\pi$-bonding $t_{\pi}$ orbital-overlap integrals for the direct electron hopping, and $t_{id}$ hopping 
amplitude from the indirect orbital overlap via the intermediate oxygen $p$-orbitals. 
We neglect higher-order processes involving the hopping between 
two different oxygen $p$-orbitals. 
To illustrate the resulting tight binding model, 
the hopping Hamiltonian between the sublattices 4 and 5 is written as follows.
\begin{eqnarray}
{\mathcal H}_{hop}^{45} & = & 
\sum_{\alpha} t_{id} (d^{\dagger}_{i,xz,\alpha}  d^{\phantom{\dagger}}_{j,yz,\alpha}  
+ d^{\dagger}_{i,yz,\alpha}  d^{\phantom{\dagger}}_{j,xz,\alpha} )
\nonumber \\
&+& \frac{t_{\pi} }{2} (d^{\dagger}_{i,xz,\alpha}-d^{\dagger}_{i,yz,\alpha} ) 
(d^{\phantom{\dagger}}_{j,xz,\alpha} - d^{\phantom{\dagger}}_{j,yz,\alpha}  )
\nonumber \\
&-&t_{\sigma} d^{\dagger}_{i,xy,\alpha} d^{\phantom{\dagger}}_{j,xy,\alpha}  + h.c.
\end{eqnarray}
The rest of the model can be written using space group symmetry operations.
In this paper, we set $t_{\pi}/t_{\sigma} = 0.1$ and $t_{id}/t_{\sigma} = 0.6$ that are 
close to the ones used in the band structure calculation in Ref.~\onlinecite{PhysRevB.81.024428}.

For the interaction part, we adopt the standard multi-orbital interactions that include
intra-orbital repulsion ($U$), inter-orbital repulsion ($U'$), Hund's coupling ($J$) 
and pairing hopping ($J'$), 
\begin{eqnarray}
{\mathcal H}_I &=& U \sum_{i,m}  n_{im\uparrow} n_{im\downarrow} 
+ \frac{U'}{2} \sum_{i, m\neq m'} n_{im}n_{im'} \nonumber \\
&+& \frac{J}{2} \sum_{i,m\neq m',\sigma\sigma'} 
d^{\dagger}_{im\sigma} d^{\dagger}_{im'\sigma'} d^{\phantom{\dagger}}_{im \sigma'} 
d^{\phantom{\dagger}}_{im'\sigma} \nonumber \\
&+& \frac{J'}{2} \sum_{i,m\neq m'} d^{\dagger}_{im\uparrow}d^{\dagger}_{im\downarrow}
d^{\phantom{\dagger}}_{im'\downarrow} d^{\phantom{\dagger}}_{im'\uparrow} . 
\label{eq:eq2}
\end{eqnarray}
Here $n_{im\alpha} = d^{\dagger}_{im\alpha} d^{\phantom{\dagger}}_{im\alpha}$ and
$n_{im} = \sum_{\alpha}  d^{\dagger}_{im\alpha} d^{\phantom{\dagger}}_{im\alpha} $. 
In the atomic limit, these four Kanamori's parameters in Eq.~\eqref{eq:eq2} 
have the relation $U = U'+J+J', J=J'$, which is assumed in the following discussions. 
Such multi-orbital interactions are in general very complicated and difficult to deal with. 
To make a progress, we follow the treatment in Ref.~\onlinecite{PhysRevB.83.134515} 
and decompose these multi-orbital interactions into the charge part ${\mathcal H}_c$ 
and the spin-orbital part ${\mathcal H}_{ex}$ with ${\mathcal H}_I = {\mathcal H}_c + {\mathcal H}_{ex}$,
where
\begin{eqnarray}
{\mathcal H}_c      &=& \frac{U}{2} \sum_i (n_i-5)^2 \ , \nonumber \\
{\mathcal H}_{ex}  &=& - J \sum_{i,m\neq m'} n_{im}n_{im'} +
\frac{J}{2} \sum_{i,m\neq m'} d^{\dagger}_{im\uparrow}d^{\dagger}_{im\downarrow}
d^{\phantom{\dagger}}_{im'\downarrow} d^{\phantom{\dagger}}_{im'\uparrow} 
\nonumber \\
&+&
 \frac{J}{2}  \sum_{i,m\neq m',\sigma\sigma'} 
d^{\dagger}_{im\sigma} d^{\dagger}_{im'\sigma'} d^{\phantom{\dagger}}_{im \sigma'} d^{\phantom{\dagger}}_{im'\sigma} .
\end{eqnarray}
Here we assume that the average electron occupation per site is 5, appropriate for 
the Ir$^{4+}$ ion with 5d$^5$ electron configuration. We neglect an unimportant 
constant in ${\mathcal H}_c$. ${\mathcal H}_c$ is the usual 
Hubbard-$U$ interaction and describes the energy cost for the electron charge fluctuation. 
${\mathcal H}_{ex}$ describes how the electrons arrange themselves among different 
spin-orbital states, i.e. on-site spin-orbital exchange interaction. 
Since ${\mathcal H}_c$ and ${\mathcal H}_{ex}$ describe two different physics, we
will treat them separately. 

\section{Slave Rotor Mean-Field Theory and Metal-Insulator Transition}
We will first consider the charge part of the interactions as
the Hubbard $U$ is the largest interaction energy scale. 
We consider the Hamiltonian $\tilde{\mathcal H} = {\mathcal H}_0 + 
{\mathcal H}_c$ and analyze the phase diagram within the slave-rotor mean-field theory\cite{PhysRevB.70.035114}. 
In the slave-rotor formalism, we decompose the electron operator into a bosonic 
charge rotor $e^{i\theta_i}$ and a fermionic spinon $f_{im\alpha}$ (that carries spin and orbital quantum numbers), {\it i.e.}
$d^{\phantom{\dagger}}_{im\alpha} \equiv e^{- i\theta_{i} } f_{im\alpha} $. 
With this decomposition, the physical Hilbert space is enlarged and we need to
impose the constraint $L_i = \sum_{m\alpha} f^{\dagger}_{im\alpha} f^{\phantom{\dagger}}_{im\alpha} -5$
to get back to the physical Hilbert space. Here $L_i$ is the angular momentum operator
conjugate to $\theta_i$, {\it i.e.} $[\theta_i, L_j ] = i \delta_{ij}$. The Hamiltonian $\tilde{\mathcal H}$ expressed
in the rotor and spinon variables is further decomposed into two mean-field 
Hamiltonians for the rotors and the spinons, respectively,
\begin{eqnarray}
{\mathcal H}_{r} &=& \frac{U}{2} \sum_{i} (L^2_i + h L_i) 
+ Q_r \sum_{\langle ij \rangle} e^{i(\theta_i - \theta_j)} 
\\
{\mathcal H}_{f}  &=& 
\sum_{i,mn,\alpha\beta} \frac{\lambda}{2}   {\bf L}_{mn} \cdot \boldsymbol{\sigma}_{\alpha\beta} 
 f^{\dagger}_{im\alpha} f^{\phantom{\dagger}}_{in\beta} 
+ \sum_{i,m,\alpha} D_{im} f^{\dagger}_{im\alpha} f^{\phantom{\dagger}}_{im\alpha}
\nonumber \\
&+& Q_f \sum_{\langle ij \rangle,  mn, \alpha} 
t_{ij,mn} f^{\dagger}_{im\alpha} f^{\phantom{\dagger}}_{jn\alpha} 
- h\sum_{i,m,\sigma} f^{\dagger}_{im\sigma} f_{im\sigma} .
\end{eqnarray}
Here $h$ is the Langrange multiplier introduced to implement the constraint on average, 
and the mean field parameters $Q_r$ and $Q_f$ are given by 
$Q_r \equiv  \sum_{mn\sigma} t_{ij,mn} 
\langle  f^{\dagger}_{im\sigma} f^{\phantom{\dagger}}_{jn\sigma}   \rangle_f$ and
$Q_f \equiv \langle  e^{i(\theta_i - \theta_j)} \rangle_{r}$, where the 
subindices indicate the mean field ground states taken to evaluate the
expectation values.
Here we have made a uniform mean field approximation such that $Q_f$ and $Q_r$  
are uniform on all the bonds. When the rotor is condensed with $\langle e^{i\theta_i} \rangle \neq 0$, 
the spinon binds with the charge rotor, form an electron with a finite quasi-particle
weight $Z\neq 0$, and the system is in a Fermi liquid phase. 
When the rotor is uncondensed with $\langle e^{i\theta_i} \rangle = 0$, 
the spinons are deconfined, form
spinon Fermi surfaces, and the system is in a U(1) QSL phase. 
We solve this mean-field Hamiltonians self-consistely.
The rotor condensation occurs when the lowest rotor
mode becomes gapless. 

\begin{figure}[htp]
\includegraphics[width=7cm]{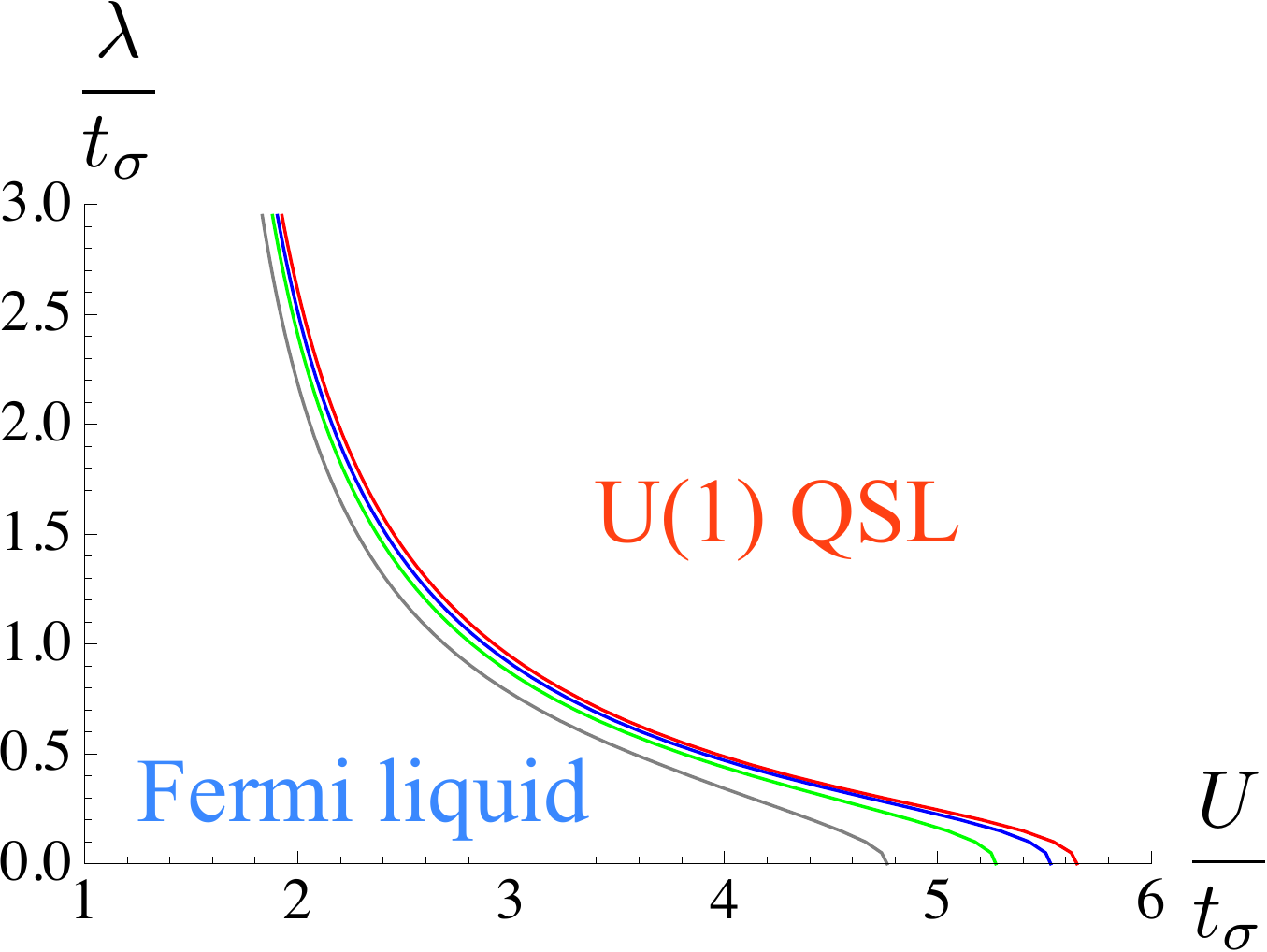}
\caption{(Color online) The slave-rotor mean-field phase diagram\cite{note} with the Hubbard-$U$ 
in the charge sector only. Different curves correspond to
different single-ion anisotropy parameters $D$. From left to right, 
$D = 0.8 t_{\sigma}, 0.4 t_{\sigma}, 0.2 t_{\sigma}, 0$. }
\label{fig:fig1}
\end{figure}

The resulting phase diagram is depicted in Fig.~\ref{fig:fig1}.
There are three different energy scales: SOC, interaction and bandwidth. 
The SOC narrows the electron bandwidth, which effectively enhances the correlation 
effect and drives a MIT at a reduced critical interaction strength. This is the key underlying reason 
for the presence of strong correlation physics in 5$d$ electron systems\cite{PesinBalents},
that were previously believed to be weakly correlated. Moreover, the 
correlation effect also suppresses the bandwidth and thus enhances the SOC effect. 

\begin{figure}[htp]
\centering
\subfigure{\includegraphics[width=8.5cm]{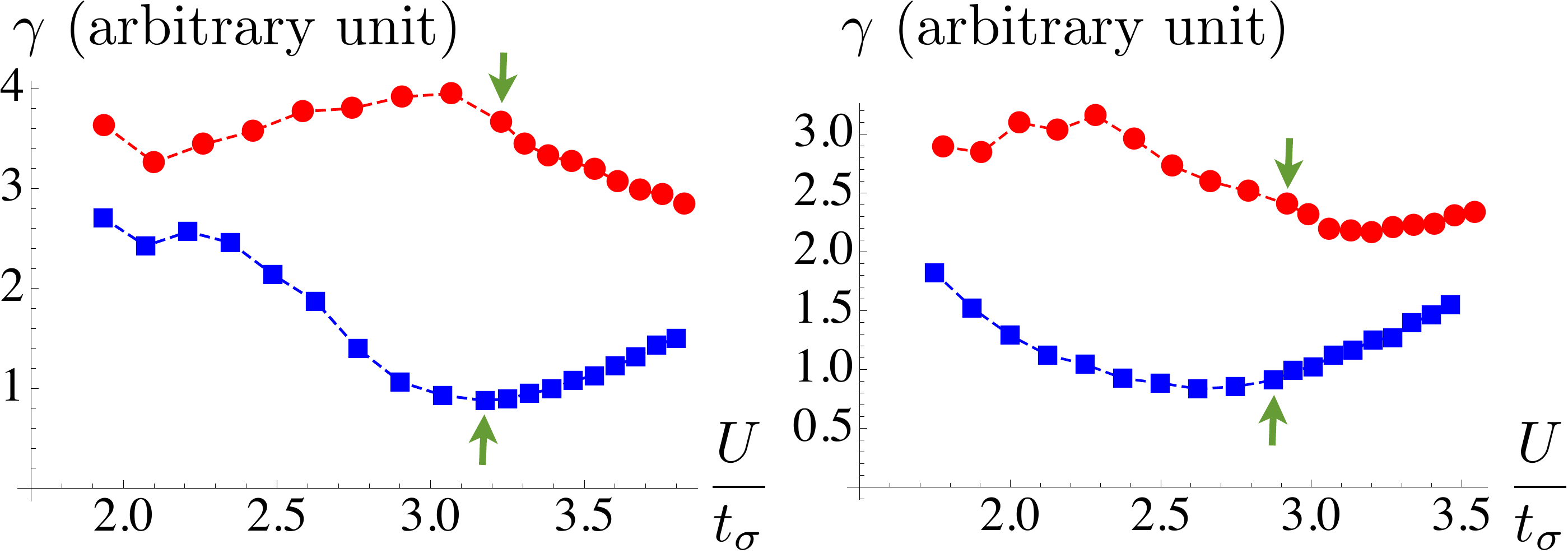}
\label{fig:fig2a}}
\subfigure{\includegraphics[width=8.5cm]{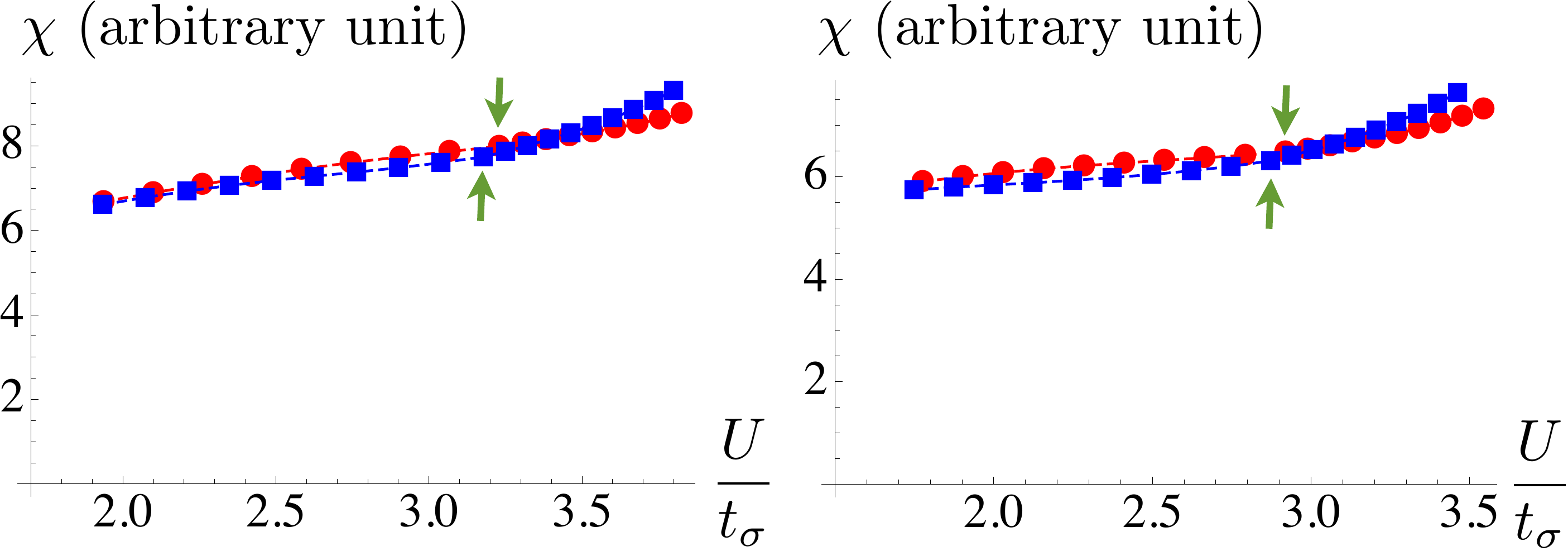}
\label{fig:fig2b}}
\subfigure{\includegraphics[width=8.5cm]{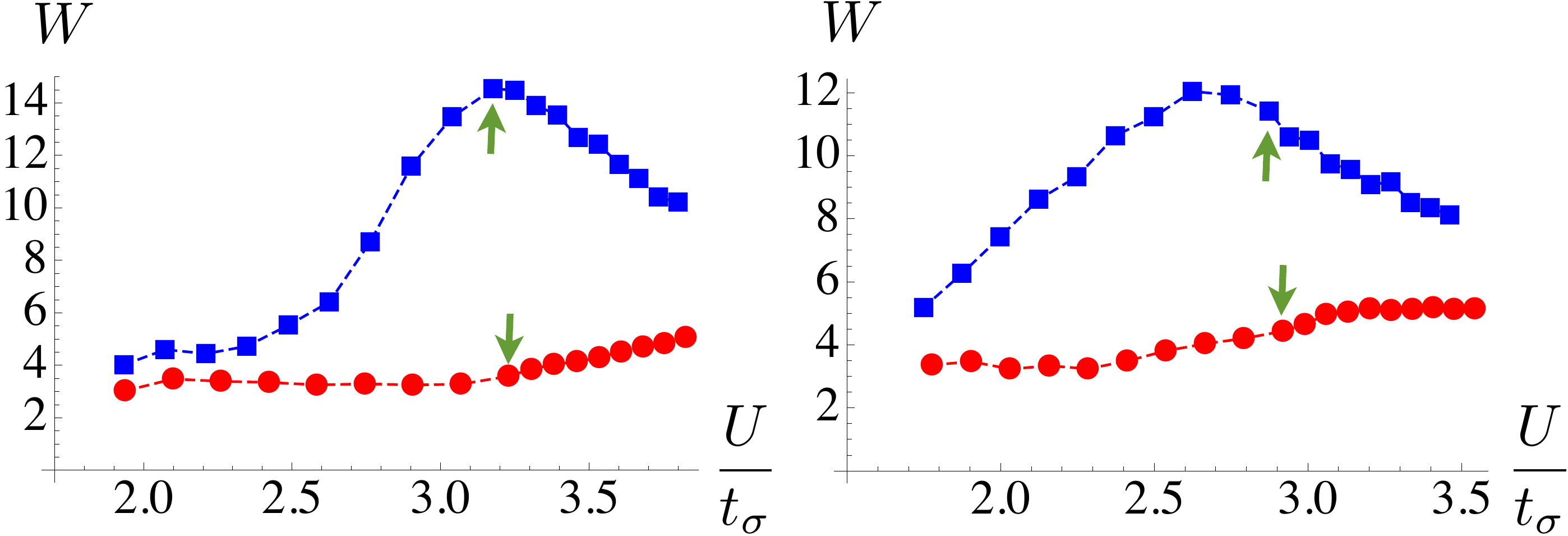}
\label{fig:fig2c}}
\caption{(Color online) The specific heat, magnetic susceptibility and Wilson ratio 
obtained from the slave-rotor mean-field theory with the Hubbard-$U$ in the charge sector only.
In the left (right) three plots, we have $\lambda = 0.8t_{\sigma}$ ($\lambda = t_{\sigma}$). 
In all the plots, the red circular (blue squared) dots are the data points with $D = 0$ ($D=0.2t_{\sigma}$). 
Dashed curves connecting the dots are the guides to the eye. 
The arrows indicate the locations of the metal-insulator transition.
The results are obtained for a finite size system with $40\times40\times 40$ unit cells at $T=0.0005t_{\sigma}$. }
\label{fig:fig2}
\end{figure}

Using the Ioffe and Larkin's relation\cite{PhysRevB.39.8988}, the heat capacity is given by
$C_v = C_{vf}+ C_{vr}$ where $C_{vf}$ ($C_{vr}$) is the spinon (rotor) contribution.  
Since the rotor contribution
is subdominant at low temperatures as $C_{vr} \propto T^3$ in the metallic phase
and $C_{vr} \propto e^{-\Delta/T}$ ($\Delta$ is the charge gap) in the Mott insulator,
the heat capacity is mostly determined by the spinon contribution $C_{v} \approx C_{vf}$. 
The magnetic susceptibility comes only from the
spinon contribution, {\it i.e.} $\chi = \chi_f$. 
In the slave-rotor mean-field theory for the Hamiltonian $\tilde{\mathcal H}$, 
the spinons are essentially treated as free fermions. In Fig.~\ref{fig:fig2},
we have computed the thermodynamic quantities for two different SOCs ($\lambda = 0.8t_{\sigma}, t_{\sigma}$) with
$D=0$ and $D= 0.2 t_{\sigma}$. 
The specific heat is not monotonic because the spinon/electron Fermi surface changes due to
the presence of the anisotropy and the SOC.
The specific heat with $D=0$ only varies slightly near the MIT, which may be consistent with 
the experiments. Taking $t_{\sigma} = 0.64 $eV and $\lambda = t_{\sigma} = 0.64 $eV\cite{PhysRevB.81.024428},
we find $\gamma \approx 3.63$mJ/molK$^2$, only slight larger than the experimental value
 ($\approx$3mJ/molK$^2$)\cite{Perry}.
The specific heat with $D=0.2t_{\sigma}$ is strongly suppressed near
the MIT in contrast to the experiments. This suggests
the actual material may not have a strong anistropy due to the IrO$_6$ distortion. 
In Fig.~\ref{fig:fig2}, we only find a small enhancement of magnetic susceptibility, 
which is inconsistent with the experiments. 
Since the Wilson ratio is sensitive to magnetic fluctuations and, moreover, it is the 
magnetic susceptibility that varies significantly in the experiments, we naturally turn to the 
on-site spin-orbital exchange ${\mathcal H}_{ex}$ that has not been so far included in our analysis.

\section{Inclusion of Spin-Orbital Exchange and Large Wilson Ratio}
We now consider the on-site spin-orbital interactions.
We follow Ref.~\onlinecite{PhysRevB.83.134515} and include the on-site exchange interaction 
${\mathcal H}_{ex}$ in the spinon mean-field Hamiltonian ${\mathcal H}_f$. We analyze its effect 
by a weak coupling analysis as the Hubbard-$U$ of the charge
sector is supposed to be a much bigger and dominant interaction.
The spinon mean-field Hamiltonian with this additional interaction
${\tilde {\mathcal H}}_{f}$ is then given by
${\tilde {\mathcal H}}_{f} = \mathcal{H}_f + {\mathcal H}_{ex}$ with ${\mathcal H}_{ex}$
written in terms of the spinon operators. 
Since the ground state does not have a magnetic order in the slave-rotor mean-field theory, 
the inclusion of the spin-orbital exchange interaction in the spinon Hamiltonian 
does not modify the mean-field phase diagram in the weak coupling approach. 
Moreover, the density of state on the Fermi surface is not modified by the inclusion of 
the spin-orbital interaction, so the heat capacity stays the same as the case without this
interaction.

\begin{figure}[htp]
\centering
\includegraphics[width=8.5cm]{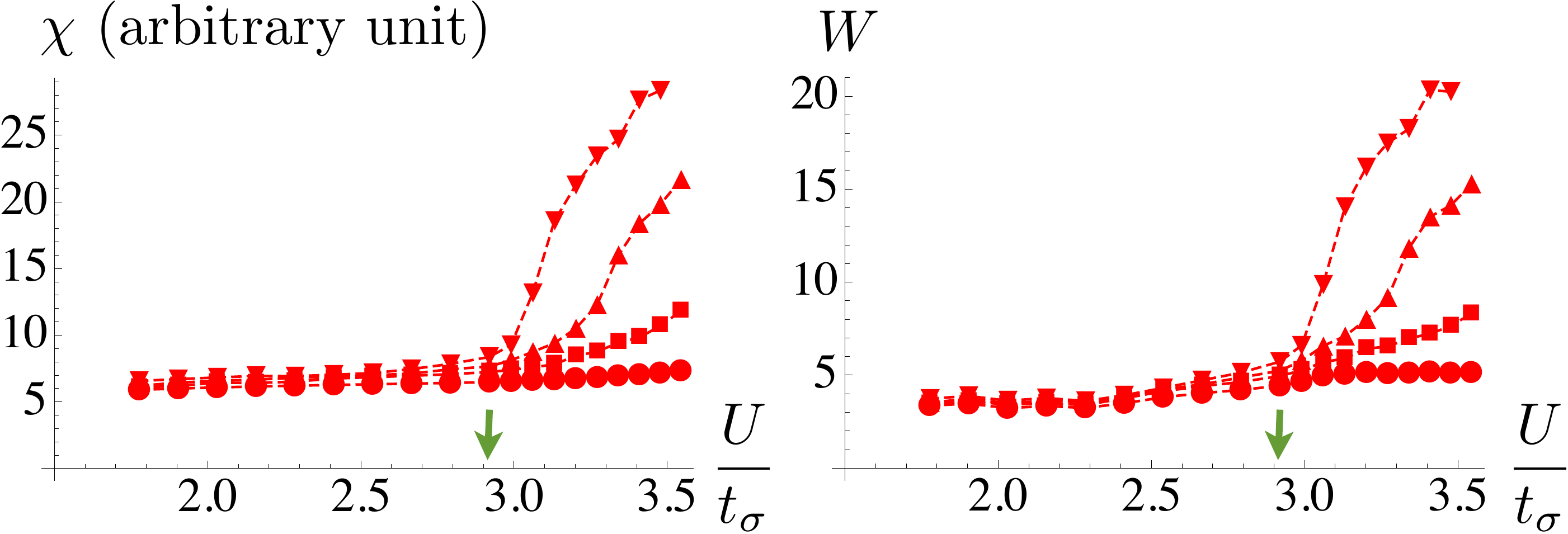}
\caption{(Color online) Magnetic susceptibilities and Wilson ratios for $\lambda = t_{\sigma}, D=0$,
when the spin-orbital exchange interaction is taken into account .
In both plots, from the top to bottom, $J=0.3t_{\sigma}, 0.2t_{\sigma}, 0.1t_{\sigma},0 $. 
Dashed curves connecting the dots are the guides to the eye. 
The arrows indicate the locations of the metal-insulator transition.
The results are obtained for a finite size system 
with $40\times40\times 40$ unit cells at $T=0.0005t_{\sigma}$.}
\label{fig:fig3}
\end{figure}

We note that the magnetic susceptibility can be strongly enhanced by this on-site 
spin-orbital exchange interaction.  
Because the spin-rotational symmetry is explicitly broken by the SOC and the orbital angular 
momentum also contributes to the magnetization, the magnetic susceptibility is not just 
determined by the density of state at the Fermi level but is sensitive to the nature of 
the whole many-body state.
This is quite different from a normal Fermi liquid with the spin-rotational invariance. 
The on-site spin-orbital exchange interaction encourages the electrons to occupy different orbitals and thus 
increases the orbital angular momentum. On the other hand, putting the electrons into a single orbital state
tends to quench the orbital angular momentum. 
The orbital angular momentum further couples to the spin degrees of freedom via the SOC. 
The presence of the spin-orbital exchange interaction, therefore, strongly
modifies the response of the system to external magnetic fields.

The resulting magnetic susceptibilities and Wilson ratios for 
different strengths of the spin-orbital exchange interaction 
are plotted in Fig.~\ref{fig:fig3}.
The magnetic susceptibility is only slightly enhanced by the spin-orbital 
exchange interaction in the metallic side, while it is strongly enhanced 
in the QSL side. Here two relevant energy scales are the electron bandwidth 
and the spin-orbital exchange interaction.
In the metallic phase, the bandwidth is rather big and a weak spin-orbital exchange 
interaction does not cause much change. 
In the QSL phase, however, the renormalized bandwidth is suppressed by the correlation effect (Hubbard-$U$)
and the SOC, which effectively enhances the on-site spin-orbital exchange interaction. 
The enhanced on-site spin-orbital exchange interaction, together with the SOC, leads 
to the large enhancement of the magnetic susceptibility and Wilson ratio. 

\section{Summary and Discussion}

In this work, we provide an explanation of the unusually large Wilson ratio 
observed in \nio\,, a candidate material for a three-dimensional
quantum spin liquid \cite{PhysRevLett.99.137207}. 
We study the behaviors of the magnetic susceptibility 
and heat capacity across the transition from the metallic state to a U(1)
quantum spin liquid with Fermi surfaces using an extended Hubbard model
that includes the spin-orbit coupling and multi-orbital interactions.
It is shown that the combination of the strong spin-orbit
coupling and spin-orbital interactions leads to much enhanced magnetic 
susceptibility in the underlying U(1) spin liquid compared to 
the metallic phase, where the effect of the multi-orbital interactions
is much weaker. On the other hand, the heat capacity does not acquire such 
enhancements in the spin liquid phase and is basically the same as that of the
metallic phase. This leads to the anomalously large Wilson ratio in
the spin liquid phase. 

Our results immediately suggest that the Wilson ratio in the metallic 
phase, that may be obtained by reducing the correlation effect, 
should be much smaller compared to the spin liquid phase
even though it is generally bigger than unity because of the 
spin-orbit coupling. As mentioned in the main text, a recent 
experiment on single crystals of \nio\, obtained both metallic
and insulating samples \cite{Perry}. While the origin of such behaviors is
not clear at the moment, thermodynamic properties of
the metallic samples are consistent with our predictions.
Namely, the magnetic susceptibility is much bigger in the
insulating samples while the heat capacity seems to be
more or less the same in both of the metallic and insulating samples.
A cleaner experiment would be to drive the transition from
the insulating quantum spin liquid phase to a metallic phase
by applying a hydrostatic pressure and measure the change
in the magnetic susceptibility and heat capacity.

The direct measurement of the spinon excitations in the spin liquid 
phase may be done by measuring the spinon particle-hole continuum in the 
inelastic spin structure factor.
While the neutron scattering on these samples is challenging, the resonant 
inelastic X-ray scattering (RIXS) may be able to observe such excitations\cite{Kim2009}.
Polarized neutron scattering may also provide information about gauge field fluctuations\cite{Lee2012}.
Thermal conductivity measurement may also provide an indirect evidence for 
the spinon Fermi surface\cite{thermal} albeit it may be small because
of the very small semi-metallic density of states at the Fermi level in \nio\,,
as seen in the small heat capacity coefficient $\gamma = C_v/T$.

\acknowledgements

We thank L. Balents, P. A. Lee and D. Podolsky for helpful discussions.
We expecially thank R. Perry and H. Takagi for sharing their experimental data before publishing. 
GC was supported by DOE award no. DE-SC0003910. YBK was supported by the NSERC, 
CIFAR, and Centre for Quantum Materials at the University of Toronto.  
Some of this work was carried out at the Kavli Institute for Theoretical Physics;
our stays there were supported by NSF grant no. PHY11-25915.

\bibliography{refs}

\begin{thebibliography}{10}
\bibitem{note}%
  \BibitemOpen
  \bibinfo {note} {We follow Ref.15 and double the critical correlation $U_c$
  at MIT to keep the right atomic limit.}%
  \bibAnnoteFile{Stop}{note}%
\end{thebibliography}

\end{document}